\PassOptionsToPackage{obeyspaces}{url}

\documentclass[a4paper,UKenglish,cleveref]{lipics-v2021}

\nolinenumbers 

\hideLIPIcs

\usepackage{booktabs,longtable}
\usepackage{xurl}
\usepackage{tikz}
\usepackage{xspace}
\usetikzlibrary{positioning, arrows.meta, calc}

\hypersetup{colorlinks=true,linkcolor=blue,citecolor=blue,urlcolor=blue}
\hypersetup{bookmarksnumbered=true}
\makeatletter
\AtBeginDocument{\let\@oddfoot\@empty\let\@evenfoot\@empty}
\makeatother

\newcommand{\doi}[1]{\href{https://doi.org/#1}{\nolinkurl{doi:#1}}}
\newcommand{\appref}[1]{\hyperref[#1]{Appendix~\ref*{#1}}}
\newcommand{\thmref}[1]{\hyperref[#1]{\mbox{Theorem~\ref*{#1}}}}
\newcommand{\figref}[1]{\hyperref[#1]{Fig.~\ref*{#1}}}
\newcommand{\tabref}[1]{\hyperref[#1]{Table~\ref*{#1}}}
\newcommand{\eqnref}[1]{\hyperref[#1]{\textup{(\ref*{#1})}}}

\newenvironment{interfaceparams}{%
  \begin{list}{}{\setlength{\leftmargin}{1em}%
    \setlength{\itemsep}{2pt}\setlength{\parsep}{0pt}%
    \setlength{\topsep}{3pt}}%
}{\end{list}}

\DeclareRobustCommand{\lean}[1]{\begingroup\def\UrlBreaks{\do\.\do\/\do\_\do\-}\def\UrlBigBreaks{}\nolinkurl{#1}\endgroup}
\DeclareRobustCommand{\toolname}[1]{\texttt{#1}}
\newcommand{\LeanPEP}{\toolname{Lean-PEP}\xspace}
\newcommand{\Lean}{\toolname{Lean}\xspace}
\newcommand{\mathlib}{\toolname{mathlib}\xspace}

\newcommand{\R}{\mathbb R}

\newcommand{\ip}[2]{\langle #1,#2\rangle}

\DeclareMathOperator*{\argmin}{argmin}
\DeclareMathOperator{\obj}{\texttt{obj}}

\crefformat{section}{#2\S#1#3}   

\title{\LeanPEP: 
Performance-Estimation Certificates to 
Formally Verified Algorithmic~Bounds
}
\titlerunning{Lean-PEP: Verified Algorithmic Bounds}

\author{Xinyuan Guo\footnote{%
Part of this work formed Xinyuan Guo's Part III project for the 2025--2026 academic year at the School of Electronics and Computer Science, University of Southampton, UK, supervised by Andersen Ang.%
}}
{Department of Earth Science and Engineering, Imperial College London, UK}
{xinyuan.guo26@imperial.ac.uk}{}{}
\author{Andersen Ang}
{School of Electronics and Computer Science, University of Southampton, UK}
{andersen.ang@soton.ac.uk}{}{}
\authorrunning{X. Guo and A. Ang}

\ccsdesc[500]{Theory of computation~Logic and verification}
\keywords{Formal verification, Lean, Certificate checking, Convex optimization, Semidefinite programming, Performance estimation, Proximal gradient}

\AtBeginDocument{%
  \hypersetup{
    pdftitle={Lean-PEP: Performance-Estimation Certificates to Formally Verified Algorithmic Bounds},
    pdfauthor={Xinyuan Guo; Andersen Ang},
    pdfsubject={Performance-estimation certificates and formal verification of algorithmic bounds},
    pdfkeywords={Formal verification, Lean, Certificate checking, Convex optimization, Semidefinite programming, Performance estimation, Proximal gradient}
  }%
}

\begin{document}
\maketitle

\begin{abstract}
Performance-estimation certificates describe worst-case bounds for optimization algorithms. 
Turning a certificate into a formal theorem requires exact verification of its algebra and a connection between its coordinates, constraints and objective and the intended function-level problem. 
Lean-PEP provides a reusable interface for this task. It checks a submitted source and exact certificate against a supported canonical finite model, and a parameterized analytic theorem transfers the resulting finite bound to every admissible algorithm trajectory. 
Generated computation proofs discharge the obligations that depend on each certificate. 
An additional capture entry checks the numerical record and reconstruction policy. 
The implemented two-step proximal-gradient consumer verifies a new input with normalized bound 62500/499999, and a fixed eight-configuration batch yields seven distinct payloads with accepted certificates and executed Lean proofs.
PG/GD cases and one-step attainment witnesses establish the corresponding algorithmic conclusions, and a fresh installation reproduces consumption of the supplied input. 
These results show how new certificate data can reuse a proved analytic interface to obtain function-level bounds within the supported models.
\end{abstract}

\section{Introduction}\label{sec:intro}
Performance estimation provides certificates of worst-case bounds for optimization algorithms.
A researcher who obtains such a certificate would like to turn it into a theorem about every function and algorithm trajectory satisfying the intended assumptions.
The challenge is to verify the certificate's exact algebra and connect that algebra to the algorithm and function class, without writing a new analytic proof for each input.
\LeanPEP makes this connection reusable: generated checks discharge certificate-dependent obligations, and a parameterized theorem supplies the function-level conclusion.

A performance estimation problem (PEP) encodes sampled function values, oracle information and algorithm steps in a finite optimization problem. This problem is often a semidefinite program assembled from interpolation inequalities~\cite{drori2014,taylor2017smooth,taylor2017composite}.
Its dual certificates express a performance bound as a nonnegative combination of constraints and a positive-semidefinite residual.
Exact arithmetic can check this identity.
To obtain the intended algorithmic claim, however, the coordinates, constraints and objective in the identity must represent the problem being analyzed.
Its assumptions and objective determine its meaning.

We organize this connection around a common finite model.
On one side, source checks bind a submitted representation to an algorithm profile, and exact certificate checks establish a bound for every feasible datum in that model.
On the other side, a reusable analytic proof maps a positive-radius trajectory to a feasible datum while preserving its normalized performance.
Applying the finite bound and restoring units gives the function-level result; the theorem handles zero radius separately through optimality.
The \emph{source} describes the submitted samples, coordinates, constraints and objective; the \emph{certificate} supplies the exact algebraic data that prove the bound.
These objects change with the input, while the proof connecting the canonical finite model to actual trajectories remains fixed.

\LeanPEP implements this design in \Lean4 with \mathlib~\cite{lean2021,mathlib2020}.
The semantic interface accepts a source and exact certificate.
A second interface, the capture interface, also checks the relation of the source and certificate to a numerical record and a declared reconstruction policy.
Both use the same analytic theorem.
For accepted inputs, a fixed renderer generates the data and computation proofs needed to apply the analytic theorem. Users who apply the theorem at the function level supply the analytic and trajectory assumptions.

The paper makes three contributions within the supported models:
\begin{itemize}
\item A reusable theorem combining target binding, exact finite soundness and an actual-trajectory embedding, with a capture entry checking numerical origin and reconstruction.
\item Explicit preservation conditions for the representation changes used in that composition, including coordinate transport, row transformations and performance lifting.
\item An implemented PG2 consumer tested on a new input, on seven distinct payloads from eight fixed configurations, and in a fresh installation.
PG/GD bounds, attainment witnesses and controlled changes to sources and certificates test the interfaces.
\end{itemize}
The bounds calibrate the stated profiles and fixed construction strategy against established rates.
\cref{sec:model} defines the mathematical objects; \cref{sec:main} explains their connection; \cref{sec:algorithm} describes candidate construction and consumption; and \cref{sec:experiments} evaluates the resulting theorems and reuse.

\subsection{Running example: a two-step proximal-gradient bound}\label{sec:overview}
Consider minimizing $F=f+h$ on a complete real inner-product space.
The function $f$ is convex and has a supplied $L$-Lipschitz gradient, where $L>0$. The function $h$ is real-valued and convex.
A chosen minimizer $x_\star$ and initial point satisfy $\|x_0-x_\star\|^2\le R^2$, where $R\ge0$.
The trajectory takes two proximal-gradient (PG) steps of size $1/L$ under \cref{sec:contract}.

The input consists of a source for this two-step problem and an exact rational certificate. The certificate contains a bound, ordered constraint multipliers and a PSD residual witness.
The accepted reuse input bounds every admissible two-step PG trajectory:
\[
F(x_2)-F(x_\star) ~\le~ LR^2B, \qquad B= 62500/499999,
\]
where $B$ is a bound for the finite model with unit smoothness and unit radius. The analytic interface restores the factor $LR^2$.
The renderer proves this input's checks by computation and applies the same generic theorem used by other accepted inputs.
We revisit this example to explain the target objective, construct the exact candidate and report verification costs.

\subsection{Notation and abbreviations}
\begingroup\small
\setlength{\tabcolsep}{-5pt}
\renewcommand{\arraystretch}{1.08}
\noindent\begin{tabular}{@{}p{.20\linewidth}p{.79\linewidth}@{}}
\multicolumn{2}{@{}l}{\textbf{Mathematical notation}}\\\midrule
$E$ & Complete real inner-product space.\\
$f,h,F\!=\!f\!+\!h$ & Smooth convex term, real-valued convex term, composite objective.\\
$x_j,\ x_0$ & Iterate at step $j$ and initial point.\\
$x_\star$ & Chosen global minimizer of $F$ for PG, or of $f$ for GD.\\
$g,\ g_j,\ g_\star$ & Supplied gradient of $f$; $g_j=g(x_j)$ and $g_\star=g(x_\star)$.\\
$s_\star,\ s_1,\ldots,s_T$ & Sampled subgradients of $h$ at $x_\star,x_1,\ldots,x_T$; $s_\star=-g_\star$.\\
$L,\ R$ & Gradient Lipschitz upper bound $L>0$; initial-distance bound $R\ge0$.\\
$T,\ n\!=\!T\!-\!1,\ k$ & Step count, horizon offset and retained-row count.\\
$y,\ G$ & Sampled values and Gram matrix.\\
$\obj(y,G), m(d)$ & Submitted finite objective and base performance metric.\\
$\tau$ & Lifted performance scalar: $\tau\le m(d)$, with equality in the trajectory lift.\\
$b,\ B$ & Normalized certificate upper bound; physical bound $LR^2B$.\\
$\lambda_i,\ S$ & Nonnegative row multipliers and dual residual matrix.\\
$M,\ \eta,\ D$ & Reconstruction denominator cap, direct error budget and grid-seed denominator.
\\
\midrule
\multicolumn{2}{@{}l}{\textbf{Abbreviations}}\\\midrule
PEP & Performance estimation problem.\\
PG, GD & Proximal gradient, gradient descent.\\
SDP & Semidefinite program.\\
PSD & Positive semidefinite.
\end{tabular}
\endgroup

\subsection{Related work}\label{sec:related}
For a researcher with a new PEP certificate, the central question is what must be supplied or checked for that input, and which mathematical connections can be reused.
PEP models express algorithmic performance through finite interpolation constraints and dual certificates~\cite{drori2014,taylor2017smooth,taylor2017composite}; \toolname{PEPit} provides a modelling and numerical solution framework~\cite{pepit2024}.
The certificate itself can become a further optimization variable.
Berg Thomsen et al.~\cite{bergthomsen2026finding} optimize over PEP dual certificates for a chosen performance target, using sparsification and semidefinite programs to search for simpler patterns of active hypotheses and residual terms.
Their procedures also recover intermediate lemmas, including Lyapunov arguments for proximal-point methods.
The resulting inequalities and residual decompositions provide concrete ingredients for subsequent analytic arguments and formal verification.
\toolname{AutoLyap}~\cite{upadhyaya2026autolyap} organizes a problem class, algorithm and chosen Lyapunov analysis into semidefinite conditions, then solves the resulting SDP to obtain Lyapunov data for the corresponding convergence analysis.

Certificate structure also guides discovery workflows.
\toolname{PEPFlow}~\cite{suh2025pepflow} supports primal and dual PEP modelling and solution, inspection of dual variables and certificate structure, and analytic proof construction with \toolname{SymPy} checks of symbolic identities.
\toolname{AutoOPT}~\cite{autoopt2026} combines PEP-based numerical method design, analytic proof discovery and \Lean formalization of selected convergence arguments, with statement comparison and author confirmation.
\toolname{Peppy}~\cite{peppy2026} organizes \toolname{PEPFlow} certificate search and Lyapunov synthesis around numerical and symbolic checks, retaining step, initial and terminal identities in inspectable notebooks.
It also provides an auxiliary \lean{lean-from-ipynb} route to \Lean formalization.
These workflows expose candidate inequalities and proof structures; accepting one as a formal input requires a checked connection to the mathematical statement it is intended to establish.

Foundational certificate checkers provide reusable machinery for this acceptance step.
Harrison~\cite{harrison2007} reconstructs rational SOS witnesses from numerical SDP output, checks positive semidefiniteness exactly and proves the resulting polynomial identity in \toolname{HOL Light}.
\toolname{ValidSDP}~\cite{validsdp2017} proves polynomial positivity in \toolname{Coq} using certified computations: coefficient-error bounds and a shifted-matrix condition verified by floating-point Cholesky with rounding-error guarantees.
The current \LeanPEP checker instead consumes exact rational coefficient identities and PSD factor identities.
In each case, checking and its soundness argument can be reused for new witness data.
For PEP, the checked coordinates, complete ordered constraints and objective must additionally denote the intended algorithm problem, so the algebraic proposition has the required meaning.

Jeong et al.~\cite[\S 5.2--5.4]{jeong2026flag} formalize flag algebras in \Lean, linking computable representations to abstract flag-algebra objects through adequacy theorems.
Reusable certificate verification and proof generators turn certificate-supported inequalities into asymptotic graph-density bounds.
For PEP, the corresponding domain obligations concern interpolation constraints, submitted coordinates and ordered rows, Gram data and actual trajectories, and the terminal objective with its physical scale restored.

Verified optimization semantics and first-order analysis supply complementary connections.
\toolname{CvxLean}~\cite{cvxlean2023} proves reductions between optimization problems, composing verified atoms and solution maps for DCP transformations and supporting custom reductions with proofs.
The input is a problem expression or transformation, and existing reduction results support its interpretation.
Li et al.~\cite{li2024} formalize gradients, subgradients, proximal operators and parameterized convergence theorems in \Lean; their LASSO example instantiates the generic proximal-gradient result with a concrete function and update contract.
These reusable results determine which functions, operators and runs an algorithmic theorem covers.

\LeanPEP establishes a PEP-specific chain through a common canonical finite model.
Source acceptance connects the submitted coordinates, complete ordered constraints and performance objective to that model.
Exact finite soundness gives an upper bound for every feasible datum in the model.
The analytic construction maps a positive-radius trajectory satisfying the function and run contract to such a datum, preserving normalized performance.
Applying the finite bound and restoring scale yields the function-level algorithmic conclusion; the same interface also handles zero radius through optimality.
\cref{sec:model} and \cref{sec:main} give the mathematical objects and proof obligations for this composition.
\tabref{tab:related-work} compares the input-dependent work and reuse units of the directly related approaches.

\begin{table}[!ht]
\centering
\begingroup
\setlength{\tabcolsep}{1pt}
\newcommand{\rwcell}[1]{{\fontsize{8pt}{8.5pt}\selectfont\raggedright #1\par}}
\begin{tabular}{@{}p{.16\linewidth}p{.285\linewidth}p{.27\linewidth}p{.235\linewidth}@{}}
\rwcell{Work} & \rwcell{Established result} & \rwcell{Input-dependent work} & \rwcell{Reusable component} \\
\hline\hline
\rwcell{Harrison~\cite{harrison2007}} &
\rwcell{Rational SOS witness to a \toolname{HOL Light} polynomial proof} &
\rwcell{Polynomial goal \& reconstructed SOS witness} &
\rwcell{Polynomial proof \& SOS verification machinery} \\
\addlinespace
\rwcell{\toolname{ValidSDP}~\cite{validsdp2017}} &
\rwcell{Numerical SOS witness to a \toolname{Coq} positivity proof} &
\rwcell{Polynomial, Gram witness \& certified error/matrix checks} &
\rwcell{Certified arithmetic, matrix verification \& reflection theorem} \\
\addlinespace
\rwcell{Jeong et al.~\cite{jeong2026flag}} &
\rwcell{Certificate-supported asymptotic graph-density bounds} &
\rwcell{Target \& forbidden graphs, exact certificate \& its checks} &
\rwcell{Computable representations, proof generators \& graph-limit semantics} \\
\addlinespace
\rwcell{\toolname{CvxLean}~\cite{cvxlean2023}} &
\rwcell{Verified reductions between optimization problems} &
\rwcell{Problem expression; proofs for new atoms or custom reductions} &
\rwcell{Verified atoms, reduction constructions \& solution transport} \\
\addlinespace
\rwcell{Li et al.~\cite{li2024}} &
\rwcell{Function \& algorithm contracts to convergence theorems} &
\rwcell{Concrete function/update instance \& contract proofs} &
\rwcell{Convex analysis, proximal theory \& convergence results} \\
\addlinespace
\rwcell{Berg Thomsen et al.~\cite{bergthomsen2026finding}} &
\rwcell{Simplified PEP certificates \& intermediate lemmas} &
\rwcell{PEP formulation, target guarantee \& certificate search} &
\rwcell{Sparsification procedures \& extracted proof patterns} \\
\addlinespace
\rwcell{\LeanPEP (fixed PG2)} &
\rwcell{Accepted certificate to a bound for admissible PG trajectories} &
\rwcell{Numerical record, exact data \& generated acceptance proofs} &
\rwcell{Validators, representation transport, analytic proofs \& renderer} 
\end{tabular}
\endgroup
\caption{Related work by its mathematical input and reuse unit. Each row concerns the objects named in its input column; the final row describes the demonstrated fixed PG2 workflow.}
\label{tab:related-work}
\end{table}

In the demonstrated fixed PG2 workflow, a new numerical record supplies its own certificate data: an exact candidate, ordered multipliers, PSD factors and associated input records; some resulting values may coincide across inputs.
The renderer generates that input's computation proofs and concrete function-level theorem, which \Lean then checks.
The two-step profile, coordinate and row conventions, source validators, representation-transport proofs and renderer remain fixed, while each input's source and trace data are checked under those rules.
Finite soundness, trajectory construction, objective correspondence, normalization and the function-level theorem are reused.
Thus per-input construction and computed acceptance supply new data to the established connection between the canonical finite model and the PG function/run contract.

\section{Functions, finite models and certificates}\label{sec:model}
\subsection{Actual functions and proximal-gradient runs}\label{sec:contract}
Let $E$ be a complete real inner-product space, $L:\R_{\ge0}$ with $L>0$, and $R\in\R$ with $R\ge0$.
Let $f:E\to\R$ be convex with supplied gradient $g:E\to E$ satisfying $\|g(x)-g(y)\|\le L\|x-y\|$ for every $x,y$.
Here $L$ is an admissible Lipschitz bound in forward-backward analysis~\cite[Problem 1.1]{combettes2005}.
Let $h:E\to\R$ be convex on all of $E$, with $x_\star$ a global minimizer of $F=f+h$.

An \emph{actual trajectory} with positive step count $T$ is a supplied run $x:\mathbb N\to E$ satisfying the proximal condition below for each $0\le j<T$, together with the initial radius condition:
\begin{equation}\label{eq:prox}
 x_{j+1}~\in~\argmin_{z\in E}
 \left\{h(z)+\frac L2
 \left\|z-\bigl(x_j-L^{-1}g(x_j)\bigr)\right\|^2\right\},
 \qquad \|x_0-x_\star\|^2\le R^2.
\end{equation}
Each step combines a gradient step with proximal minimization~\cite[\S 4.2]{parikh2014}. The existence of the displayed minimizers is part of the supplied-run contract.
Moreau~\cite[\S 3]{moreau1965} develops proximity operators in Hilbert space, and Combettes and Wajs~\cite[\S 2.1 and 2.3]{combettes2005} give the subgradient inequality and proximal characterization used here.

Composite optimality supplies the reference subgradient $-g_\star$ of $h$, where $g_\star=g(x_\star)$ may be nonzero.
The finite profile samples $h$ at $x_\star,x_1,\ldots,x_T$, without requiring a subgradient at $x_0$.
Composite stationarity permits nonzero component oracles whose sum vanishes.

\subsection{Finite data and exact upper certificates}
The canonical PG model has a positive horizon $T=n+1$ and unit parameters $L=R=1$.
All symbols in this finite definition denote normalized data in centered coordinates.
For $R>0$, the analytic proof maps a trajectory in physical units into this model and later rescales its bound by $LR^2$. The same interface handles zero radius separately.
The finite definition uses the following component sample sets and reference relations:
\[
 I_f=\{\star,0,\ldots,T\},\qquad I_h=\{\star,1,\ldots,T\},\qquad
 x_\star=0,\qquad s_\star=-g_\star.
\]
The point expressions encode unit steps while allowing a nonzero reference gradient $g_\star$:
\begin{equation}\label{eq:pg-points}
 x_{j+1}=x_j-g_j-s_{j+1}\quad(0\le j<T),\qquad
 x_j=x_0-\sum_{\ell=0}^{j-1}(g_\ell+s_{\ell+1})\quad(0\le j\le T).
\end{equation}
Thus the profile represents the updates through coordinates rather than storing separate update rows.
It uses the vector family and value order
\[
 v=(x_0,g_\star,g_0,\ldots,g_T,s_1,\ldots,s_T),\qquad
 y=(f_\star,f_0,\ldots,f_T,h_\star,h_1,\ldots,h_T).
\]
Both have $2T+3$ entries. Inner products and squared norms are linear in the Gram matrix $G_{ab}=\ip{v_a}{v_b}$.
Feasibility imposes every ordered interpolation row
\begin{equation}\label{eq:interp}
\begin{aligned}
 q^f_{ij}&=f_j-f_i+\ip{g_j}{x_i-x_j}+\tfrac12\|g_i-g_j\|^2\le0
 &&(i,j\in I_f),\\
 q^h_{ij}&=h_j-h_i+\ip{s_j}{x_i-x_j}\le0
 &&(i,j\in I_h),
\end{aligned}
\end{equation}
together with $\|x_0\|^2\le1$ and $G\succeq0$.
The factor $\tfrac{1}{2}$ expresses normalized smoothness.
Taylor et al.~\cite[Corollary 1]{taylor2017smooth} characterize finite Euclidean interpolation and treat component oracles separately in composite PEP~\cite[\S 2--3]{taylor2017composite}.
The profile stores the negatives of these rows as explicit nonnegative constraints for both components; $h$ has no oracle sample at $x_0$.

The canonical \lean{objectiveQ} selects the final composite difference from the value slots:
\begin{equation}\label{eq:pg-objective}
 \obj_T=(f_T+h_T)-(f_\star+h_\star).
\end{equation}
Its evaluation on normalized actual data is $(F(x_T)-F(x_\star))/(LR^2)$ for $R>0$, as established by the objective and scaling proofs in \cref{sec:main}.
The base model uses this objective directly. The extra performance coordinate $\tau$ in a submitted source is handled by the lift in \cref{sec:lift}.

Rows and objectives are affine in the finite values and Gram entries:
\[
 e(y,G)=e_0+e_y^\top y+\sum_{a,b}e_{ab}G_{ab}.
\]
Submitted representations may reorder values, add performance or origin coordinates, and use oriented, positively scaled rows.
The source adapters connect these rows $q_i(y,G)\le0$ and their objective $\obj(y,G)$ to the canonical representation.

An exact upper certificate supplies a rational bound $b$, ordered nonnegative multipliers $\lambda_i$, and a residual matrix $S$ with a rational PSD factorization:
\begin{equation}\label{eq:identity}
 b-\obj(y,G) ~= ~
 -\sum_i\lambda_iq_i(y,G)+\ip{S}{G},
 \qquad
S ~=~
 \sum_r w_r u_ru_r^\top,\quad w_r\ge0.
\end{equation}
The identity is checked coefficient by coefficient, including constants, sampled values and both Gram axes.
In the captured PG route all interpolation inequalities are explicit rows, and the list of implicit interpolation terms is empty.

\begin{lemma}[Finite certificate soundness]\label[lemma]{lem:finite}
If \eqnref{eq:identity} holds, $\lambda_i,w_r\ge0$, and a datum satisfies $q_i(y,G)\le0$ and $G\succeq0$, then the finite bound $\obj(y,G)\le b$ holds.
\end{lemma}
\begin{proof}
Every $-\lambda_iq_i(y,G)$ is nonnegative and $\ip{S}{G}=\sum_r w_r u_r^\top Gu_r\ge0$.
\end{proof}
This weak-duality argument, oriented here for maximization, is the algebraic basis of the upper and PSD checkers; Boyd and Vandenberghe~\cite[\S 5.1.3]{boyd2004} give the corresponding minimization convention.
Source checks bind these rows and objective to the PG model.

\subsection{Binding a source to the canonical model}\label{sec:source}
A \emph{source} is a submitted representation of the sampled problem: coordinates, value slots, ordered rows, rules, labels and objective.
A profile specifies which algorithm and performance measure these data must encode.
Source checking connects that representation to the \emph{canonical finite model}. In this common model, accepted certificates give bounds and actual trajectories give feasible data.
For the running example, the profile specifies two PG steps and the final composite difference $F(x_2)-F(x_\star)$ in normalized units.

The complete record below is from \lean{GramPEP.UpperCertificate}; only its enclosing namespace and documentation are omitted from this excerpt.
\begin{lstlisting}[mathescape=true,columns=fullflexible]
structure UpperCertQ {p m : $\mathbb N$} (model : ModelQ p m) where
    bound : $\mathbb Q$
    objective : LinearExprQ p m
    interpTerms : List (InterpTermQ p)
    assumptionTerms : List (AssumptionTermQ model)
    dualMatrix : Gram m $\mathbb Q$
    dualFactor : PSDFactorCertQ m
\end{lstlisting}
Each \lean{AssumptionTermQ model} has a rational \lean{weight}.
Its \lean{index} has type \lean{Fin model.assumptions.length}.
The dependent type restricts references to rows of this model. It also fixes the coefficient dimensions.
Rational weights can still be negative, and the certificate carries its own objective.
The theorem \lean{checkUpperCertQ_sound} supplies the sign, PSD and coefficient-identity facts needed for finite soundness.

The sparse \lean{SourceQ} separately retains the base source, row rules, coverage map, labels and objective; its \lean{model} forms assumptions from the oriented submitted rows.
Thus local validity alone says nothing about whether those rows describe PG.
The joint \lean{checkBoundQ} additionally checks the source, objective equality and transport. It also checks that \lean{interpTerms} is empty.
Its theorem \lean{checked_certificate} produces a certificate for \lean{Profiles.ProximalGradient.modelQ n}. The certificate has an accepted upper check, the canonical objective and the same bound. This provides finite soundness using a certificate for the intended algorithmic model.

One PG2 source row makes this correspondence concrete.
The captured row named \lean{IC_h_convexity(x_star, x2)} has logical label \lean{convexPair} $(0,2)$ in the nonsmooth sample order $(\star,1,2)$.
Under its \lean{leZero} convention it represents
\begin{equation}\label{eq:source-convex-row}
 q^h_{\star,2}=h_2-h_\star-\ip{s_2}{x_2-x_\star}\le0.
\end{equation}
Its value coefficients select $h_2-h_\star$; its Gram coefficients expand the inner product using the native expression for $x_2$.
Checked coordinate and value transport connects this row to the negative of the canonical nonsmooth interpolation expression.
For the actual normalized trajectory $z$, the second proximal step supplies its endpoint subgradient:
\[
 \widehat s_2=z_1-\widehat g(z_1)-z_2\in\partial\widehat h(z_2),
 \qquad
 \widehat h(0)\ge\widehat h(z_2)+\ip{\widehat s_2}{0-z_2}.
\]
Evaluating the transported row on \lean{actualData} gives precisely this subgradient inequality.
The construction separately proves Gram PSD and all ordered smooth, nonsmooth and radius constraints, yielding a feasible datum for finite soundness.

The supported source transformations account for native-to-centered coordinates, storage permutations, oriented and positively scaled rows, a performance-variable lift and sparse row coverage.
Retained references remain ordered, including those with a zero multiplier.
Sparse coverage permits omission only when every row coefficient is zero before performance and origin coordinates are eliminated. Elimination cannot hide missing nonzero rows.

Coordinate changes illustrate why the directions of these checks matter.
If $C$ maps native vectors to canonical vectors, primal Gram transport is
\[
G_{\rm can} ~=~ CG_{\rm nat}C^\top.
\]
For a section $J$ with $CJ=I$, the following pairing identity holds:
\[
\ip{S_{\rm nat}}{JG_{\rm can}J^\top} ~=~ \ip{J^\top S_{\rm nat}J}{G_{\rm can}}.
\]
The transported dual residual is therefore $J^\top S_{\rm nat}J$.
The origin convention and round-trip checks justify the section on represented data; they do not require $JC=I$ on all native coordinates.
Together, the representation checks preserve the bound, coefficient identity, nonnegative terms and PSD pairing needed for finite soundness.

\subsection{Supported mathematical and executable inputs}\label{sec:support}
The semantic theorem is parameterized by every positive horizon $T=n+1$ and retained-row count $k$. Its acceptance condition checks the particular source and certificate. The executable consumer implements a fixed two-step profile within that mathematical interface. \tabref{tab:support} summarizes the mathematical and executable layers.

\begin{table}[t]\centering\small
\begin{tabular}{p{.14\linewidth}p{.82\linewidth}}
Layer & Supported contract and scope
\\\hline\hline
Semantic \quad theorem & Every positive horizon $T=n+1$ and retained-row count $k$, conditional on source and certificate acceptance; the complete real inner-product space and function/run contract of \cref{sec:contract}.\\
Representation checks & Specified native/centered coordinate maps and origin convention; storage permutations; signed rows with positive scales; one performance lift; exact zero-row coverage. Each transformation is checked.\\
Capture trace & The trace checker requires \lean{leZero} rules with scale $1$. It maps value slots; Gram axes and prepared-row order stay fixed. Raw binding retains all ordered references and zeros.\\
Executable consumer & \lean{convert(...,2)} and the PG2 validator fix 8 native vectors, 20 labels/references and a 158-entry reconstruction policy. Recovery scans 64 prescribed grids.\\
Executed cases & Separate PG1/2/3 and GD1/2 upper cases and two one-step exact pairs; relocatable consumption, fresh-install consumption and the reported batch concern PG2.
\end{tabular}
\caption{Support of the mathematical interfaces, representation checks and executable consumer.}\label{tab:support}
\end{table}

Positive row rescaling is checked at the representation layer. The captured command uses the unit-scale convention stated in the table. The next section first explains how an exact certificate is consumed, then adds the trace and numerical policy needed by that command.

\section{From a certificate to a function-level bound}\label{sec:main}
The main interface takes a source for a supported PG profile and a certificate over that source.
Its acceptance condition checks both the certificate's finite algebra and the source's relation to the canonical model.
Once accepted, the certificate gives a bound for every actual trajectory satisfying \cref{sec:contract}.
One analytic proof covers all accepted certificates.

In \Lean, the submitted source is represented by \lean{SourceQ n k}, and \lean{UpperCertQ s.model} attaches the certificate to that source's model.
The Boolean \lean{checkBoundQ s c} checks the source, local upper certificate, empty implicit-interpolation list, objective agreement and transported certificate.
These checks provide the premise of the semantic theorem.

\begin{theorem}[Semantic certificate consumption]\label{thm:semantic}
Let $n,k\in\mathbb N$, $T=n+1$, $s:\mathtt{SourceQ}\ n\ k$ and $c:\mathtt{UpperCertQ}\ s.\mathtt{model}$.
If this source and certificate satisfy
\[
\mathtt{checkBoundQ}\ s\ c ~=~ \mathtt{true},
\]
then, for every $E,L,R,f,h,g,x_\star,x$ satisfying the complete contract of \cref{sec:contract},
\begin{equation}\label{eq:actual}
 f(x_T)+h(x_T)-\bigl(f(x_\star)+h(x_\star)\bigr)
 ~\le~ (L:\R)R^2(c.\mathtt{bound}:\R).
\end{equation}
\end{theorem}
The full parameters, alternative capture signature and declaration correspondence appear in \appref{app:interface}.
For the running example, the accepted bound is 62500/499999 and $T=2$.
New certificates supply data and acceptance proofs to the theorem establishing \eqnref{eq:actual}.

\subsection{Where the certificate and trajectory meet}
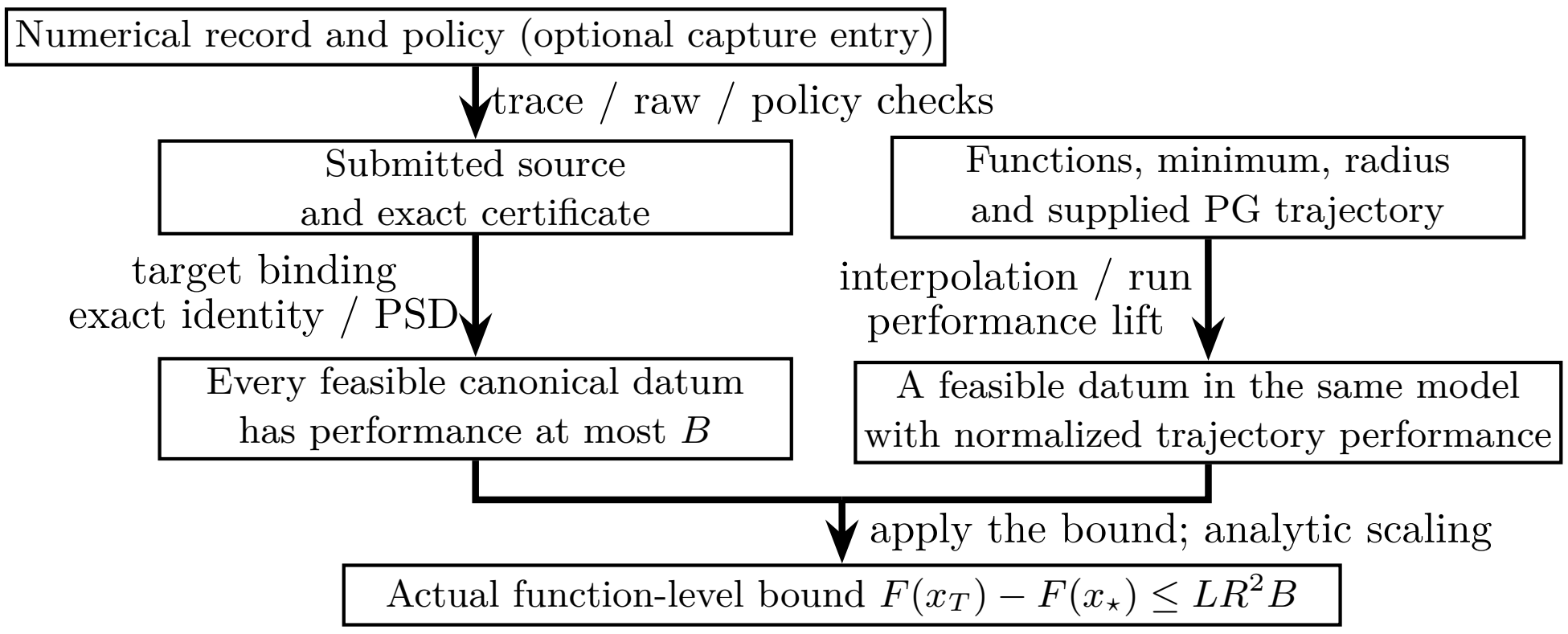
\begin{figure}[htbp]
\centering
\small
\begin{tikzpicture}[
    scale=1.75,
    >=Stealth,
    box/.style={
        draw,
        thick,
        rectangle,
        align=center,
        font=\small,
        inner xsep=2pt,
        inner ysep=2pt,
        minimum width=52mm
    },
    note/.style={
        font=\scriptsize,
        align=center,
        inner sep=0pt
    },
    arrow/.style={->, ultra thick}
]

\node[box] (record)
    {Numerical record and policy
     (optional capture entry)};

\node[box, below=6mm of record] (source)
    {Submitted source\\
     and exact certificate};

\node[box, right=8mm of source] (functions)
    {Functions, minimum, radius\\
     and supplied PG trajectory};

\node[box, below=10mm of source] (upperbound)
    {Every feasible canonical datum\\
     has performance at most $B$};

\node[box, below=10mm of functions] (witness)
    {A feasible datum in the same model\\
     with normalized trajectory performance};

\coordinate (merge) at
    ($(upperbound.south)!0.5!(witness.south)+(0,-1.75mm)$);

\node[
    box,
    minimum width=82mm,
    anchor=north
] (result) at ($(merge)+(0,-3mm)$)
    {Actual function-level bound 
     $F(x_T)-F(x_\star)\le LR^2B$};

\draw[arrow] (record.south) -- (source.north)
    node[note, midway, right=1mm]
    {\normalsize trace / raw / policy checks};

\draw[arrow] (source.south) -- (upperbound.north)
    node[note, midway, left=1mm]
    {\normalsize target binding\\ \normalsize
     exact identity / PSD};

\draw[arrow] (functions.south) -- (witness.north)
    node[note, midway, left=1mm]
    {\normalsize interpolation / run\\
     \normalsize performance lift};

\draw[ultra thick] (upperbound.south) |- (merge);
\draw[ultra thick] (witness.south) |- (merge);

\draw[arrow] (merge) -- (result.north)
    node[note, midway, right=2mm]
    {\normalsize apply the bound; analytic scaling};

\end{tikzpicture}

\caption{The proof branches meet in the same canonical finite model.
The actual-to-finite arrow preserves normalized performance;
analytic scaling handles zero radius separately.}
\label{fig:semantics}
\end{figure}

\figref{fig:semantics} shows the two branches of the proof.
The certificate establishes a universal finite bound.
The analytic contract supplies one feasible datum to which that bound applies.

\begin{proof}[Composition of the implemented proofs]
Source acceptance transports the local certificate to the canonical PG model.
Sparse coverage restores certified zero rows. Performance substitution, row normalization, ordering and coordinate transport preserve the certificate's bound and its algebraic identity.
Exact finite soundness applies the transported certificate to bound the objective of every feasible datum in that canonical model.

For $R>0$, \lean{SmoothScaling.normalizedFunction} and \lean{normalizedGradient} normalize the supplied functions and trajectory as follows:
\[
z_j=\frac{x_j-x_\star}{R},\qquad \widehat f(z)=\frac{f(x_\star+Rz)-f(x_\star)}{LR^2},\qquad \widehat g(z)=\frac{g(x_\star+Rz)}{LR},
\]
with the same normalization for $h$.
The lemma \lean{isSmoothProxRun_iff_normalized} transports each supplied proximal minimizer in physical units to a unit step. The lemma \lean{normalizedPoint_radius} gives the initial squared-distance bound in unit coordinates.

The finite vector family contains the initial displacement, reference gradient, sampled gradients and step residuals.
In unit coordinates, each residual is $\widehat s_{j+1}=z_j-\widehat g(z_j)-z_{j+1}$.
The proximal characterization proves that this residual is a subgradient of $\widehat h$ at $z_{j+1}$. The lemma \lean{position_realization} then proves by induction that the symbolic updates denote the actual points.
The construction \lean{actualData} pairs the Gram matrix of these vectors with the sampled values. The Gram matrix is PSD because it is constructed from inner products.
The proof \lean{actualData_constraints} applies smooth interpolation and convex subgradient inequalities to every ordered sample pair. It then evaluates the radius row.
The reference oracle obtained from \lean{subgradient_of_composite_minimizer} is $-\widehat g(0)$. The component oracles sum to zero at the reference point while each may be nonzero.

The lemma \lean{objective_eval} identifies the canonical objective with the composite gap.
Together with \lean{normalized_objective}, this lemma expresses the canonical objective as $(F(x_T)-F(x_\star))/(LR^2)$.
The finite bound at this datum, rescaled by $LR^2>0$, yields \eqnref{eq:actual}.

For $R=0$, the radius hypothesis forces $x_0=x_\star$.
The lemma \lean{stationary_run} uses the reference subgradient and uniqueness of the supplied proximal minimizer at each step to prove $x_j=x_\star$ throughout the run.
Both sides of \eqnref{eq:actual} are then zero, for any accepted bound.
This branch requires no division by the radius.
\end{proof}

\begin{samepage}
\subsection{Why performance lifting preserves the bound}\label{sec:lift}
A PEP source may express the performance objective through an extra scalar $\tau$.
This gives the solver a scalar objective even when the quantity of interest is an affine expression in sampled data.
For the two-step example, this expression is the normalized final composite difference.
A row requires $\tau$ to be no greater than this difference.
Feasible base data admit feasible extensions with equality, transferring universal bounds on $\tau$ to performance.
\end{samepage}

Let $d$ denote the base value/Gram data, $m(d)$ the base performance metric, and $\operatorname{extend}(d,t)$ the datum with a new scalar slot $\tau=t$ and the same Gram matrix.
Under the checked lift and a strictly positive performance scale,
\begin{equation}\label{eq:lift}
 \mathrm{RowsFeasible}\bigl(\operatorname{extend}(d,t)\bigr)
 ~~~~ \Longleftrightarrow ~~~~
 \mathrm{BaseRowsFeasible}(d)\ \land\ t\le m(d).
\end{equation}
This is a pointwise equivalence of row constraints.
In the internal nonnegative convention, the performance row is a positive multiple of $m(d)-t$.

Adding the condition that the Gram matrix is PSD to both sides gives the corresponding equivalence of universal finite bounds.
If $m(d)\le B$ for every feasible base datum, then any feasible extension satisfies $t\le m(d)\le B$.
Conversely, a universal bound on feasible lifted scalars applies to the choice $t=m(d)$ in \eqnref{eq:lift}, proving $m(d)\le B$.

At certificate level, exact substitution replaces $\tau$ by the base objective.
The declared performance row becomes the zero polynomial, and all other ordered references are preserved.
This finite argument precedes the analytic construction: only afterward does a normalized actual trajectory supply $d$.
The declarations \lean{ProximalGradientPerformance.rows_feasible_iff} and \lean{finite_upper_equivalence} implement the two finite equivalences. The scaling lemmas above connect their base datum to the actual function contract.

\subsection{Adding a numerical record and reconstruction policy}\label{sec:capture}
The capture interface is useful when an exact certificate is reconstructed from a numerical endpoint.
Each finite decimal in the endpoint is interpreted as an exact rational; these coefficients form the raw certificate data.
A trace connects the source's ordered rows and rules to the observations.
Raw binding checks that the exact raw bound, objective, multipliers and residual match that trace.
A declared policy then compares the final candidate directly with the raw certificate data. It does not accumulate a budget across intermediate candidates.

For the PG2 consumer the policy covers 158 entries and requires
\begin{equation}\label{eq:policy}
 M=10^6,\qquad \eta=10^{-6},\qquad
 \operatorname{den}(c_j)\le M,\qquad |c_j-\mathrm{raw}_j|\le\eta .
\end{equation}
Here $c_j$ and $\mathrm{raw}_j$ denote corresponding candidate and raw coefficients.
References and list lengths must agree.
The separate factor policy checks the exact PSD identity and nonnegative weights. Its factor coefficients are excluded from the coefficient list subject to the denominator and error constraints.
The semantic and capture checks jointly bind the exact certificate to its numerical record under the declared reconstruction policy.

\begin{theorem}[Capture-policy entry]\label{thm:main}
With the notation and contract of \thmref{thm:semantic}, let $t:\mathtt{TraceQ}\ n\ k$, $p:\mathtt{PolicyQ}$ and $\mathrm{raw}:\mathtt{UpperCertQ}\ s.\mathtt{model}$.
If
\begin{equation}\label{eq:hc}
 \mathtt{checkCapturedQ}\ s\ t\ p\ \mathrm{raw}\ c~~=~~\mathtt{true},
\end{equation}
then every admissible trajectory satisfies the function-level bound \eqnref{eq:actual}.
\end{theorem}
Here \lean{checkCapturedQ} checks the trace, raw binding and reconstruction obligations; its success implies acceptance by the semantic interface.
An exact target-valid certificate can also be consumed directly by \thmref{thm:semantic}, without these numerical-origin requirements.

\subsection{Closing the checks for a concrete input}\label{sec:generated}
The generic theorem is conditional on acceptance; generated computation proofs establish that condition for the supplied exact data.
The fixed renderer generates seven checker facts, a bound fact and a coefficient-count fact, all proved by \lean{decide +kernel}.
It then generates \lean{actual_upper}, which applies \thmref{thm:main} using the computed acceptance proof.
The running example yields $62500/499999$ under the same analytic and run contract.

This computation-proof pattern is also used in \toolname{ValidSDP}~\cite[\S 4.2]{validsdp2017}.
The \Lean4 \toolname{sos} tactic~\cite{morrison2026sos} proves \lean{Certificate.checks} by \lean{decide +kernel}, applies its soundness theorem, and connects the verified polynomial statement to the user's original expression.
\LeanPEP similarly supplies a concrete acceptance proof to an established soundness interface, here yielding the PEP function-level bound under its analytic and run contract.
Automatically generated source size, kernel execution cost and human proof editing are different quantities: the nine facts and theorem application are generated, while the generic analytic proof is reused.
The call and declaration mapping appear in \appref{app:interface}, and costs in \cref{sec:cost}.

\section{Constructing and consuming exact certificates}\label{sec:algorithm}
The numerical input supplies approximate coefficients; the function-level interface consumes an exact certificate.
The candidate constructor connects these two objects by using the source equations to enforce the certificate identity and the raw coefficients to choose its remaining degrees of freedom.
For the PG2 running example, the construction is a fixed, bounded search within the declared reconstruction policy.
Its output is accepted only after the exact checks and the generated \Lean proof succeed.

\subsection{Three steps in the source-affine construction}
\paragraph*{Solve the constant and value equations.}
The certificate identity \eqnref{eq:identity} requires every constant and value coefficient of $b-\obj+\sum_i\lambda_iq_i$ to vanish. For the supported PG2 source, these conditions form nine exact affine equations $Az=r$ in the 21 unknowns $z=(b,\lambda_0,\ldots,\lambda_{19})$. Gaussian elimination determines pivot variables from free variables, selecting pivots from left to right and the first available nonzero row. The producer rechecks consistency and the completed solution in rational arithmetic.

\paragraph*{Use raw seeds to reconstruct the residual.}
A trial chooses a denominator $D$ and forms seeds from the exact raw rationals.
The radius multiplier is rounded upward because the source's constant equation ties that multiplier to the bound $b$.
Every other raw value $v$ is seeded at $\lfloor Dv+\tfrac{1}{2}\rfloor/D$.
Variables in the free columns retain their seeds. The affine equations determine the variables in the pivot columns.
The source, bound and multipliers determine every residual coefficient:
\[\textstyle
S_{ij} ~=~ -\obj_{ij}+\sum_\ell\lambda_\ell(q_\ell)_{ij}.
\]
Both Gram axes enter the exact source identity; raw data stay fixed.

\paragraph*{Check nonnegativity, PSD and the direct reconstruction policy.}
Each completed candidate is compared directly with the original raw input under \eqnref{eq:policy}.
The current strategy first checks the coefficient policy and multiplier nonnegativity, then constructs and rechecks an exact PSD witness.
A candidate passing these tests must also pass the source and complete upper checks.
The fixed renderer emits the instance proofs and applies the generic function-level theorem; selected data are published only after \Lean succeeds.
Thus the affine solve enforces the constant/value equations, residual reconstruction enforces the Gram equations, and acceptance checks the sign, PSD and policy obligations that remain.

\begin{figure}[t]
\centering
\fbox{\begin{minipage}{.95\linewidth}
\textbf{Algorithm: The fixed PG2 source-affine consumer}
\begin{enumerate}
\item Check source, trace, raw and policy; form the constant/value system $Az=r$.
\item For $D=500000,499999,\ldots,499937$, form raw grid seeds, rounding the radius multiplier upward and all others to nearest grid.
\item Retain free seeds and solve for pivots, stopping if the source system is inconsistent.
\item Reconstruct the complete matrix residual from the source and the solved coefficients.
\item Reject trials failing direct policy, multiplier nonnegativity or exact PSD checks.
\item Generate typed data and a \toolname{Lean} Replay when source and upper checks first succeed.
\item Run \toolname{Lean}, audit its axioms and publish selected data on success; otherwise stop.
\end{enumerate}
Exhausting all 64 trials reports failure of this bounded search.
\end{minipage}}
\caption{Candidate construction supplies data to the fixed semantic interface. Source equations, direct raw policy and exact certificate checking have separate roles.}\label{fig:algorithm}
\end{figure}

The initial denominator 500000, descending search and 64-trial limit in \figref{fig:algorithm} are fixed implementation choices.
Half-integral source coefficients can introduce denominators dividing $2D$ when multipliers lie on a $1/D$ grid.
This gives a structural relation between the grid denominator and the coefficient denominators.
Acceptance checks every covered coefficient, including solved pivots, against $M=10^6$ and $\eta=10^{-6}$.
\appref{app:engineering} records the affine-system pivots and exact coefficient audit for the PG2 reuse input.

\subsection{Exact PSD witnesses and the consumer boundary}
The residual must be positive semidefinite even when it is singular.
The factor producer uses rational Schur elimination, an exact-arithmetic form of the block PSD criterion in Boyd and Vandenberghe~\cite[Appendix A.5.5]{boyd2004}.
It first rejects an asymmetric matrix.
A negative Schur diagonal entry gives a negative quadratic witness.
A zero diagonal with a nonzero remaining row gives another witness, lifted through earlier pivots to the original matrix.

Otherwise, the first positive active diagonal entry $d$ defines a vector $v$ by dividing its column by $d$.
The producer records $dvv^\top$ and subtracts it from the residual.
Repeating this step constructs a rational rank-one, or $LDL^\top$-style, factorization.
If all active diagonals are zero and the row checks pass, the remaining matrix is exactly zero.
The factors reconstruct the original matrix entry by entry with nonnegative weights, including when it is singular.

Numerical-to-rational SOS reconstruction likewise combines approximate candidates, exact affine identities and PSD factors~\cite[\S 2.2.2 and 3]{peyrl2008}.
The \Lean4 \toolname{sos} implementation~\cite{morrison2026sos} also uses rational $LDL^\top$ reconstruction and nonnegative rationally weighted squares, whose weights and polynomial identity undergo exact acceptance checks.
In \LeanPEP, the data Gram matrix $G$ represents sampled vectors, whereas the certificate residual $S$ certifies a nonnegative pairing with $G$.
Sampled values and Gram expressions retain the finite PEP structure, and source-dependent constraints support model binding.
Trajectory construction and the terminal objective connect this representation to algorithmic performance; the capture entry additionally enforces the direct raw-to-candidate policy.
The reusable command supports the PG2 profile; PG3 uses a related source-affine grid with separate resource settings.
Other recorded cases use different information: PG1 uses an ordered multiplier family and a rank-three PSD structure, whereas the GD1 exact pair uses the target $1/6$.
\appref{app:engineering} records these construction inputs and the content guard separating the generic mathematical closure from the concrete applications and their regressions.

\subsection{Instantiating the PG2 consumer}
A new PG2 record supplies the source, trace and raw coefficients.
The constructor produces \lean{structuredUpperQ}, including its exact PSD witness, while the renderer emits \lean{sourceQ}, \lean{traceQ}, \lean{policyQ} and \lean{rawUpperQ}.
The following excerpt shows a check from the accepted reuse \lean{Replay.lean}. The checker namespace and the other generated declarations are omitted.
\begin{lstlisting}[columns=fullflexible]
theorem structured_check :
       checkCapturedQ sourceQ traceQ policyQ rawUpperQ structuredUpperQ
       = true := by
    decide +kernel
\end{lstlisting}
The checker is \lean{LeanPEP.Adapters.ProximalGradientCapture.checkCapturedQ}.
The generated \lean{actual_upper} then passes \lean{structured_check} to \lean{ProximalGradientCaptureAudit.upper_analytic}, followed by the function/run hypotheses in \appref{app:interface}.
Candidate data and computation proofs change with the numerical input; trajectory realization, normalization and finite soundness remain shared.
The archived reuse Replay and exit-zero log establish this call for $62500/499999$.
The user of the function-level theorem supplies the analytic contract. The acceptance proof is already complete in the generated theorem.

\section{Bounds, target binding and reuse evidence}\label{sec:experiments}
\subsection{Algorithm-level bounds and one-step attainment}
The first question is which algorithmic conclusions the accepted certificates establish.
\tabref{tab:bounds} gives the normalized upper bounds for the PG and GD cases, including the PG2 running example.
For the PG rows, the Reference column uses
\[
 B_{\mathrm{ref}}(T)=\frac1{4T},\qquad T\ge1,
\]
from Teboulle and Vaisbourd~\cite[Theorem 7]{teboulle2023}.
The comparison concerns Euclidean composite convex minimization with an $L$-Lipschitz gradient for the smooth component, constant PG steps of size $1/L$, initial distance at most $R$ from a minimizer, and the last-iterate objective gap; unit normalization gives the displayed reference value.
Taylor et al.~\cite[\S 4, Tables 2--3]{taylor2018pg} earlier studied mixed-performance PEP bounds and their non-strongly-convex limits, including this value.
The Selected $B$ is the rational upper bound supported by the reconstructed exact certificate; $B-B_{\mathrm{ref}}$ is its slack relative to the reference, and multiplying by $LR^2$ restores that slack to the original units.
The semantic interface turns each accepted finite upper bound into a PG conclusion under the function and supplied-run contract of \cref{sec:contract}.
The GD references use the smooth-convex last-iterate bound at step $1/L$~\cite[Theorems 3.1--3.2]{drori2014}.

\begin{table}[t]\centering
\begin{tabular}{lrrl}
Case & Selected $B$ & Reference & $B-\text{reference}$ (approximately)\\
\hline\hline
PG1 & $1/4$ & $1/4$ & $0$\\
PG2, original & $62497/499974$ & $1/8$ & $1/1999896$ $(5.0002600\times 10^{-7})$\\
PG3 & $41630/499559$ & $1/12$ & $1/5994708$ $(1.6681380\times 10^{-7})$\\
GD1, original & $83333/499997$ & $1/6$ & $1/2999982$ $(3.3333533\times 10^{-7})$\\
GD2 & $50000/499999$ & $1/10$ & $1/4999990$ $(2.0000040\times 10^{-7})$\\
GD1, exact pair & $1/6$ & $1/6$ & $0$\\
PG2, new frozen input & $62500/499999$ & $1/8$ & $1/3999992$ $(2.5000050\times 10^{-7})$
\end{tabular}
\caption{Normalized upper bounds and independent attainment of the exact one-step pairs.}\label{tab:bounds}
\end{table}

Upper certificates and attainment witnesses answer different questions.
The PG1 upper certificate pairs with an external finite primal at $1/4$.
The GD1 exact dual and primal pair at $1/6$, using the same symbolic source across distinct numerical traces.
In the GD1 gap case, the upper and primal values are 83333/499997 and 83333/500000, respectively. Their exact positive gap is 249999/249998500000.
Analytic witnesses attain both exact one-step bounds.

\begin{corollary}[One-step actual bounds and attainment]\label[corollary]{cor:pairs}
Under the PG contract in \cref{sec:contract} with $T=1$, the certificate gives $F(x_1)-F(x_\star)\le LR^2/4$.
Under the analogous smooth GD contract with $x_1=x_0-L^{-1}g(x_0)$ and a global minimizer of $f$, the exact certificate gives $f(x_1)-f(x_\star)\le LR^2/6$.
For every $L>0,R\ge0$, each value is attained by an instance on $\R$.
\end{corollary}
The following PG instance on $\R$ attains its upper bound:
\[
f(z)=Lz,\quad g(z)=L,\quad h(z)=L\bigg(\frac{R}{2}\max\{z,0\}-z\bigg),\quad x_\star=0,\ x_0=R,\ x_1=\frac{R}{2}.
\]
Its reference gradient $g_\star=L$ is nonzero.
For GD, set $a=R/3$ and take $f(z)=Lz^2/2$ for $z\le a$, $f(z)=L(az-\tfrac{a^2}{2})$ for $z\ge a$, and $g(z)=L\min\{z,a\}$. The initial point is $R$.
This one-sided quadratic--linear witness has a quadratic left branch; the classical symmetric Huber construction has two linear tails~\cite[\S 3.1, Huber example]{parikh2014}.
The analytic witness proofs verify the full contracts, including $R=0$.
Consequently, the finite primal data establish finite attainment, while these functions establish attainment of the actual one-step bounds.

\subsection{Why the target model must be checked}\label{sec:controls}
Exact certificate algebra proves a statement about its submitted assumptions.
To obtain the intended PG claim, those assumptions must also match the target model.
The forged-source example illustrates this distinction by replacing one assumption with $-\obj$ under \Lean's nonnegative-row convention. The submitted model therefore assumes $\obj\le0$.
A certificate with bound zero, weight one on this row and zero residual passes the local upper check.
The joint target check rejects the source.
The forged assumption makes the local certificate algebra valid while changing the PG problem.

A valid synchronized transformation provides the positive comparison.
At the Rows layer, alternating row directions and positive scales $2$, $1/3$, $5/2$, $7$ and $4/5$ are accompanied by matching row data and multipliers.
The checked transformation preserves the certificate expression and its bound.
This example establishes transport under the Rows layer's stated conditions; the captured CLI retains its unit-scale trace convention.
The exact control declarations, including \lean{forgedSourceQ}, \lean{forged_local_passes} and \lean{forged_joint_rejected}, appear with the synchronized example in \appref{app:engineering}.

Three further controls separate the roles of the checks.
A looser certificate that is valid for the target model has a proved function-level upper bound. However, it exceeds the raw reconstruction budget.
A small residual perturbation passes the source and complete policy checks but fails the exact identity check.
Doubling rows and halving multipliers without synchronizing the rules and trace preserves the local certificate but fails representation binding.
This last transformation preserves the feasible domain, whereas the forged source changes an assumption.
Together the controls show why target binding, exact algebra and numerical policy are distinct acceptance obligations.

\subsection{Reusing the interface for a new PG2 input}\label{sec:reuse}
The running example tests whether a fixed consumer can derive a new function-level theorem from a different numerical endpoint.
The generic source and producer were frozen before a preplanned \toolname{Clarabel} run with regularization $10^{-7}$, absolute/relative gap and feasibility tolerances $10^{-10}$, 300 iterations and equilibration enabled.
The endpoint had status \lean{optimal_inaccurate}/\lean{AlmostSolved} and differed from the three distinct payloads in forty historical records.
Source, coordinate orders and references passed PG2 validation.

The fixed strategy accepted its second grid, produced six exact PSD factors, and obtained
\[
B
~=~
62500/499999
~=~
1/8 + 1/3999992.
\]
The renderer automatically generated the nine instance facts and the function-level theorem using the existing analytic interface.
The direct reconstruction audit checked all 158 policy entries. The largest raw-to-candidate error was approximately $9.93186677\times10^{-7}$ at multiplier 7. The maximum denominator among the covered entries was 999998.
These checks establish consumption of this input under the frozen policy without changing the generic proof.

Differences from earlier numerical payloads and reconstruction error are measured separately.
Each of the three earlier payloads differs from the new raw endpoint in 85 of the 158 covered entries.
In the displayed comparison, the largest raw difference between the new and earlier endpoints occurs at zero-based multiplier index 10. The value changes from $0.6714451237581617$ to $0.65794427613641$.
That comparison concerns two solver endpoints, not the reconstruction budget for a candidate.
\appref{app:engineering} reports both exact comparisons and the associated affine-solve details for this input.

\subsection{All outcomes of a fixed PG2 batch}\label{sec:batch}
A bounded batch examines the same reuse path across several prescribed solver settings.
Eight complete configurations, their order, participating source/harness hashes and resource limits were fixed before any new output was observed.
\tabref{tab:batch} reports every configuration.
The stack was \toolname{PEPit} 0.5.1, \toolname{CVXPY} 1.8.2 and \toolname{Clarabel} 0.11.1~\cite{pepit2024,cvxpy2016,clarabel2026}, with $T=2$, $L=R=1$ and step size one.
Option parsing was checked without solving.
Solver stopping tolerances control numerical residuals and gaps~\cite[\S 2.6]{clarabel2026}; all eight configurations retain the mathematical model and all 158 entries of the exact policy \eqnref{eq:policy}.

Each configuration received one solve with a 120-second outer wall limit.
The candidate/source phase also had a 120-second outer limit, recovery retained the same 64-grid search, and every distinct accepted candidate received one sequential \Lean run with a 900-second limit.
The fixed, nonrandom configurations include deliberate reference settings.
They measure recovery across numerical payloads of the same PG2 model.

\begin{table}[t]\centering
\begin{tabular}{lccclll}
ID & Reg. & Tolerance & Eq. & Solver status & Payload relation & Recovery
\\
\hline\hline
B01 & $10^{-8}$ & $10^{-8}$ & yes & optimal & historical & accepted\\
B02 & $10^{-7}$ & $10^{-10}$ & yes & inaccurate & historical & accepted\\
B03 & $10^{-9}$ & $10^{-10}$ & yes & inaccurate & new & accepted\\
B04 & $10^{-6}$ & $10^{-10}$ & yes & optimal & new & accepted\\
B05 & $10^{-8}$ & $10^{-9}$ & yes & inaccurate & historical & accepted\\
B06 & $10^{-8}$ & $10^{-11}$ & yes & inaccurate & same as B05 & reuse B05\\
B07 & $10^{-8}$ & $10^{-10}$ & no & inaccurate & historical & accepted\\
B08 & $10^{-7}$ & $10^{-9}$ & no & optimal & new & accepted
\end{tabular}
\caption{All eight preplanned configurations. Tolerance means all three gap-absolute, gap-relative and feasibility tolerances; all use 300 iterations. ``Inaccurate'' is the recorded \toolname{CVXPY} status \texttt{optimal\_inaccurate}. Full effective settings and raw solver output are archived.}\label{tab:batch}
\end{table}

Eight solves completed and yielded seven distinct full numerical payloads.
The source, basis, value order and references were checked for agreement before deduplication. Deduplication used exact Fractions for the complete numerical object.
Four payloads matched the four distinct historical payloads represented by the 41 archived captures. The other three payloads were new (B03, B04 and B08).
B06 equalled B05. The deterministic recovery and \Lean result were reused, with no second execution.
All seven distinct payloads produced accepted candidates with six exact PSD factors and a proof of \lean{actual_upper} checked by Lean, as shown in \tabref{tab:batchbounds}.
The source/reference checks, all 158 direct-policy entries and the separate factor policy were preserved for each of the seven distinct payloads.

\begin{table}[t]\centering
\begin{tabular}{lrrrr}
Payload & Accepted grid $D$ & Grids tried & Bound $B$ & $B-1/8$
\\
\hline\hline
B01 & 499974 & 27 & $62497/499974$ & $1/1999896$\\
B02 & 499999 & 2 & $62500/499999$ & $1/3999992$\\
B03 & 499972 & 29 & $62497/499972$ & $1/999944$\\
B04 & 499990 & 11 & $62499/499990$ & $1/1999960$\\
B05 & 499999 & 2 & $62500/499999$ & $1/3999992$\\
B07 & 499998 & 3 & $31250/249999$ & $1/1999992$\\
B08 & 499983 & 18 & $62498/499983$ & $1/3999864$
\end{tabular}
\caption{Exact selected bounds for distinct payloads. B06 is associated with B05. Fractions are the producer's checked outputs; the reference $1/8$ is used only to compute the displayed excess.}\label{tab:batchbounds}
\end{table}

The recovery diagnostics describe the trials before these candidates were accepted.
Across seven recoveries, 92 grids were visited. Of these, 85 failed the direct numerical policy check. The other seven passed the nonnegativity and exact PSD checks, the \toolname{Python} identity diagnostic and \Lean verification.
Nonnegativity was recorded as true at all 92 grids.
The 85 trials that failed the policy check stopped before the PSD, identity and \Lean stages. They therefore provide no outcomes for those later stages.
No later-stage rejection, search exhaustion or timeout occurred.
The outer observer retains every rejected trial and its returned policy subchecks; unrecorded coefficient-level failure details remain unknown.
The records thus cover seven related payloads and every recovery trial in this fixed PG2 batch.

\subsection{Fresh-installation reproducibility and execution cost}\label{sec:cost}
To test reproducibility beyond the existing development environment, we executed the consumer in a fresh directory using \Lean~4.31.0 and nine pinned upstream dependencies.
Inherited \Lean/\toolname{Python} paths were cleared, the matching \mathlib cache was obtained, and the project Core was built from source.
The consumer successfully verified the archived PG2 reuse input and proved its function-level bound $B=62500/499999$ in \lean{actual_upper}.

For the seven distinct payloads in \cref{sec:batch}, candidate construction took 1.57--1.96 seconds per payload. \Lean verification took 46.46--68.46 seconds per payload.
The fresh Core build took 85.86 seconds, followed by 48.67 seconds for verification of the supplied input.
Measurements were obtained on an Apple M1 Pro with 32 GiB RAM.
These results demonstrate reproducible verification for the supplied input under the pinned dependencies and characterize the execution cost of the supported PG2 consumer.
Detailed timing, memory, profiling and environment records are retained in the artifact.

\subsection{What changes when the input structure changes}\label{sec:structural-reuse}
\tabref{tab:structural-reuse} distinguishes reusable components from the data, adapters and proofs needed when the input structure changes.
Its evidence column separates executed cases, parameterized declarations inspected in the source, and obligations for extensions to the supported models.

\begin{table}[t]\centering\small
\begin{tabular}{@{}p{.16\linewidth}p{.23\linewidth}p{.28\linewidth}p{.24\linewidth}@{}}
\raggedright Change & \raggedright Reused components & \raggedright New data or proof work & \raggedright Evidence status\tabularnewline\midrule
\raggedright PG2 numerical endpoint & \raggedright Source validators, exact checks, analytic bridge \& renderer & \raggedright Raw record, exact candidate \& factors, generated acceptance proofs & \raggedright Executed: reuse input \& seven batch payloads\tabularnewline[3pt]
\raggedright Supported representation & \raggedright Coordinate/row transport, binding \& zero-row coverage & \raggedright Coordinate maps, row rules/references \& checked synchronization & \raggedright Executed: Rows \& representation controls\tabularnewline[3pt]
\raggedright PG horizon & \raggedright Semantic interfaces for $T=n+1$ \& retained-row count $k$ & \raggedright Source dimensions; capture, renderer \& recovery adaptations; instance proofs & \raggedright Source audit; separate PG1/2/3 Replays; CLI fixed at PG2\tabularnewline[3pt]
\raggedright PG versus GD & \raggedright Rational affine/Gram certificates, PSD \& upper checks & \raggedright Algorithm profile, source binding \& actual-trajectory bridge & \raggedright Executed: PG1/2/3 \& GD1/2 cases\tabularnewline[3pt]
\raggedright Performance metric & \raggedright Exact affine identity \& PSD checking & \raggedright Target expression, source/lift binding \& actual performance equality & \raggedright Extension obligation of the final-gap interface\tabularnewline
\end{tabular}
\caption{Reuse across input changes, with evidence scoped to the supported profiles.}\label{tab:structural-reuse}
\end{table}

\subsection{A runnable code artifact}\label{sec:review-artifact}
The complete repository snapshot identified in \appref{app:artifacts} is the code artifact for this paper. It contains the Core proof modules, PG2 consumer, dependency locks and archived reuse input with its accepted certificate.
Run \lean{scripts/setup.py}, then \lean{tools/replay_pg_certificate.py verify}, using the commands in \appref{app:reproduce}.
Successful verification produces \lean{Replay.lean}, \lean{replay.log} and \lean{actual_upper} with bound $62500/499999$.
\appref{app:reproduce} gives the environment, commands and trust boundary; \appref{app:artifacts} maps historical records to the source snapshots on which they were executed.

Source code and reproduction instructions are available\footnote{Project repository and reproduction entry: \url{https://github.com/weartem13/lean-pep}.}.
The \lean{SOURCE_MANIFEST.json} manifest identifies the versioned sources and inputs.

\section{Conclusion}
\LeanPEP turns an accepted exact performance-estimation certificate into a bound for every function and algorithm trajectory satisfying the supported contract.
The certificate checks and the trajectory construction meet in the same canonical finite model.
This connection makes the analytic proof reusable: a new certificate changes the data and generated computation proofs, while the parameterized function-level theorem remains the same.
The capture interface adds checks relating the exact certificate to its numerical record and reconstruction policy.
The PG and GD cases, seven distinct payloads in the fixed PG2 batch, and fresh-installation experiment demonstrate this path from certificates to formally verified bounds.

Further mathematical work could extend certificate consumption to additional performance metrics and parameterized algorithm families.
Each extension would connect its finite model to a function-level contract through reusable algebraic checks and representation transports.
Certificate simplification and intermediate Lyapunov lemmas, as developed by Berg Thomsen et al.~\cite{bergthomsen2026finding}, provide methods for discovering potential-function structure from PEP certificates.
The systematic Lyapunov searches in \toolname{AutoLyap}~\cite{upadhyaya2026autolyap} offer a complementary source of candidate potentials for selected problem and algorithm classes.
A future \LeanPEP development could connect such local inequalities or potential functions to exact checking and actual-function semantics.
It would verify the descent relation, initial estimate and link to terminal performance, then use induction to establish parameterized performance bounds for the resulting algorithm families and metrics.

These developments could support an agent-guided optimization research loop.
Models could propose algorithm parameters, performance bounds and proof structures, and choose the next experiment; exact computation and formal verification would assess the resulting mathematical claims.
Dual certificates and independently verified function instances could guide the search for sharp rates and optimal parameter choices within specified algorithm families and parameter domains.
The semantic interface would provide a reusable basis for discovering and proving new properties of optimization algorithms.

\phantomsection
\pdfbookmark[1]{References}{references}
\bibliographystyle{unsrturl}
\bibliography{references}

@article{drori2014,
  author = {Drori, Yoel and Teboulle, Marc},
  title = {Performance of first-order methods for smooth convex minimization: a novel approach},
  journal = {Mathematical Programming},
  volume = {145}, number = {1--2}, pages = {451--482}, year = {2014},
  doi = {10.1007/s10107-013-0653-0},
  url = {https://arxiv.org/abs/1206.3209}
}

@article{taylor2017smooth,
  author = {Taylor, Adrien B. and Hendrickx, Julien M. and Glineur, Fran{\c c}ois},
  title = {Smooth strongly convex interpolation and exact worst-case performance of first-order methods},
  journal = {Mathematical Programming},
  volume = {161}, pages = {307--345}, year = {2017},
  doi = {10.1007/s10107-016-1009-3},
  url = {https://arxiv.org/abs/1502.05666}
}

@article{taylor2017composite,
  author = {Taylor, Adrien B. and Hendrickx, Julien M. and Glineur, Fran{\c c}ois},
  title = {Exact worst-case performance of first-order methods for composite convex optimization},
  journal = {SIAM Journal on Optimization},
  volume = {27}, number = {3}, pages = {1283--1313}, year = {2017},
  doi = {10.1137/16M108104X},
  url = {https://arxiv.org/abs/1512.07516}
}

@article{pepit2024,
  author = {Goujaud, Baptiste and Moucer, C{\'e}line and Glineur, Fran{\c c}ois and Hendrickx, Julien M. and Taylor, Adrien B. and Dieuleveut, Aymeric},
  title = {{PEPit}: computer-assisted worst-case analyses of first-order optimization methods in {Python}},
  journal = {Mathematical Programming Computation},
  volume = {16}, pages = {337--367}, year = {2024},
  doi = {10.1007/s12532-024-00259-7},
  url = {https://arxiv.org/abs/2201.04040}
}

@inproceedings{cvxlean2023,
  author = {Bentkamp, Alexander and Fern{\'a}ndez Mir, Ramon and Avigad, Jeremy},
  title = {Verified reductions for optimization},
  booktitle = {Tools and Algorithms for the Construction and Analysis of Systems (TACAS 2023), Part II},
  series = {Lecture Notes in Computer Science}, volume = {13994},
  pages = {74--92}, publisher = {Springer}, year = {2023},
  doi = {10.1007/978-3-031-30820-8_8},
  url = {https://arxiv.org/abs/2301.09347}
}

@misc{li2024,
  author = {Li, Chenyi and Wang, Ziyu and He, Wanyi and Wu, Yuxuan and Xu, Shengyang and Wen, Zaiwen},
  title = {Formalization of Complexity Analysis of the First-order Algorithms for Convex Optimization},
  year = {2024}, eprint = {2403.11437}, archivePrefix = {arXiv},
  url = {https://arxiv.org/abs/2403.11437v3},
  note = {{arXiv}:2403.11437, version 3 (21 July 2024); consulted 13 October 2026}
}

@misc{autoopt2026,
  author = {Kim, Heechang and Ryu, Ernest K. and Das Gupta, Shuvomoy},
  title = {A Domain-Specific Harness for End-to-End Automation of Optimization Research},
  year = {2026}, eprint = {2608.07407}, archivePrefix = {arXiv},
  url = {https://arxiv.org/abs/2608.07407v1},
  note = {{arXiv}:2608.07407, version 1 (7 August 2026); consulted 13 October 2026}
}

@misc{peppy2026,
  author = {Suh, Jaewook J. and Yoon, TaeHo and Nguyen, Edward Duc Hien and Ying, Bicheng and Ma, Shiqian},
  title = {{Peppy}: An {AI}-Assisted Workflow for Tight Convergence Analysis of Optimization Algorithms},
  year = {2026}, eprint = {2609.35762}, archivePrefix = {arXiv},
  url = {https://arxiv.org/abs/2609.35762v1},
  note = {{arXiv}:2609.35762, version 1 (28 September 2026); consulted 13 October 2026}
}

@inproceedings{harrison2007,
  author = {Harrison, John},
  title = {Verifying Nonlinear Real Formulas via Sums of Squares},
  booktitle = {Theorem Proving in Higher Order Logics (TPHOLs 2007)},
  series = {Lecture Notes in Computer Science}, volume = {4732},
  pages = {102--118}, publisher = {Springer}, year = {2007},
  url = {https://www.cl.cam.ac.uk/~jrh13/papers/sos.pdf}
}

@inproceedings{validsdp2017,
  author = {Martin-Dorel, {\'E}rik and Roux, Pierre},
  title = {A Reflexive Tactic for Polynomial Positivity using Numerical Solvers and Floating-Point Computations},
  booktitle = {Proceedings of the 6th ACM SIGPLAN Conference on Certified Programs and Proofs (CPP 2017)},
  pages = {90--99}, publisher = {ACM}, year = {2017},
  doi = {10.1145/3018610.3018622},
  url = {https://www.onera.fr/sites/default/files/447/cpp2017.pdf}
}

@book{boyd2004,
  author = {Boyd, Stephen and Vandenberghe, Lieven},
  title = {Convex Optimization},
  publisher = {Cambridge University Press}, year = {2004},
  doi = {10.1017/CBO9780511804441},
  url = {https://web.stanford.edu/~boyd/cvxbook/}
}

@article{parikh2014,
  author = {Parikh, Neal and Boyd, Stephen},
  title = {Proximal Algorithms},
  journal = {Foundations and Trends in Optimization},
  volume = {1}, number = {3}, pages = {123--231}, year = {2014},
  doi = {10.1561/2400000003},
  url = {https://web.stanford.edu/~boyd/papers/prox_algs.html}
}

@article{moreau1965,
  author = {Moreau, Jean-Jacques},
  title = {Proximit{\'e} et dualit{\'e} dans un espace hilbertien},
  journal = {Bulletin de la Soci{\'e}t{\'e} Math{\'e}matique de France},
  volume = {93}, pages = {273--299}, year = {1965},
  doi = {10.24033/bsmf.1625},
  url = {https://www.numdam.org/item/BSMF_1965__93__273_0/}
}

@article{combettes2005,
  author = {Combettes, Patrick L. and Wajs, Val{\'e}rie R.},
  title = {Signal Recovery by Proximal Forward-Backward Splitting},
  journal = {Multiscale Modeling \& Simulation},
  volume = {4}, number = {4}, pages = {1168--1200}, year = {2005},
  doi = {10.1137/050626090},
  url = {https://pcombet.math.ncsu.edu/mms1.pdf}
}

@article{taylor2018pg,
  author = {Taylor, Adrien B. and Hendrickx, Julien M. and Glineur, Fran{\c c}ois},
  title = {Exact Worst-Case Convergence Rates of the Proximal Gradient Method for Composite Convex Minimization},
  journal = {Journal of Optimization Theory and Applications},
  volume = {178}, number = {2}, pages = {455--476}, year = {2018},
  doi = {10.1007/s10957-018-1298-1},
  url = {https://arxiv.org/abs/1705.04398}
}

@article{teboulle2023,
  author = {Teboulle, Marc and Vaisbourd, Yakov},
  title = {An elementary approach to tight worst case complexity analysis of gradient based methods},
  journal = {Mathematical Programming},
  volume = {201}, number = {1--2}, pages = {63--96}, year = {2023},
  doi = {10.1007/s10107-022-01899-0},
  url = {https://link.springer.com/article/10.1007/s10107-022-01899-0}
}

@article{peyrl2008,
  author = {Peyrl, Holger and Parrilo, Pablo A.},
  title = {Computing sum of squares decompositions with rational coefficients},
  journal = {Theoretical Computer Science},
  volume = {409}, number = {2}, pages = {269--281}, year = {2008},
  doi = {10.1016/j.tcs.2008.09.025},
  url = {https://www.mit.edu/~parrilo/pubs/files/PeyrlParrilo-ComputingSumOfSquaresDecompositionsWithRationalCoefficients.pdf}
}

@inproceedings{lean2021,
  author = {de Moura, Leonardo and Ullrich, Sebastian},
  title = {The {Lean} 4 Theorem Prover and Programming Language},
  booktitle = {Automated Deduction -- CADE 28},
  series = {Lecture Notes in Computer Science}, volume = {12699},
  pages = {625--635}, publisher = {Springer}, year = {2021},
  doi = {10.1007/978-3-030-79876-5_37},
  url = {https://lean-lang.org/papers/lean4.pdf}
}

@inproceedings{mathlib2020,
  author = {{The mathlib Community}},
  title = {The {Lean} Mathematical Library},
  booktitle = {Proceedings of the 9th ACM SIGPLAN International Conference on Certified Programs and Proofs (CPP 2020)},
  publisher = {ACM}, year = {2020},
  doi = {10.1145/3372885.3373824},
  url = {https://leanprover-community.github.io/papers/mathlib-paper.pdf}
}

@article{cvxpy2016,
  author = {Diamond, Steven and Boyd, Stephen},
  title = {{CVXPY}: A {Python}-Embedded Modeling Language for Convex Optimization},
  journal = {Journal of Machine Learning Research},
  volume = {17}, number = {83}, pages = {1--5}, year = {2016},
  url = {https://jmlr.org/papers/v17/15-408.html}
}

@article{clarabel2026,
  author = {Goulart, Paul J. and Chen, Yuwen},
  title = {{Clarabel}: an interior-point solver for conic programs with quadratic objectives},
  journal = {Mathematical Programming Computation}, year = {2026},
  doi = {10.1007/s12532-026-00320-7},
  url = {https://link.springer.com/article/10.1007/s12532-026-00320-7},
  note = {Published online 21 May 2026; volume and pagination not assigned on the consulted publisher record}
}

@misc{bergthomsen2026finding,
  author = {Berg Thomsen, Daniel and Upadhyaya, Manu and Goujaud, Baptiste and Dieuleveut, Aymeric and Taylor, Adrien},
  title = {Finding Simple Proofs for First-Order Optimization},
  year = {2026}, eprint = {2607.08753}, archivePrefix = {arXiv},
  primaryClass = {math.OC},
  url = {https://arxiv.org/abs/2607.08753v1},
  note = {{arXiv}:2607.08753, version 1 (9 July 2026); consulted 6 November 2026}
}

@misc{jeong2026flag,
  author = {Jeong, Gyeongwon and Park, Seonghun and Hyun, Jihoon and Oum, Sang-il and Yang, Hongseok},
  title = {Formalizing Flag Algebras in {Lean}},
  year = {2026}, eprint = {2607.23500}, archivePrefix = {arXiv},
  primaryClass = {cs.LO},
  url = {https://arxiv.org/abs/2607.23500v2},
  note = {{arXiv}:2607.23500, version 2 (14 October 2026); consulted 10 November 2026}
}

@inproceedings{suh2025pepflow,
  author = {Suh, Jaewook J. and Ying, Bicheng and Jiang, Xin and Nguyen, Edward Duc Hien},
  title = {{PEPFlow}: A {Python} Library for the Workflow of Performance Estimation of Optimization Algorithms},
  booktitle = {{NeurIPS} 2025 Workshop: {ScaleOPT}: {GPU}-Accelerated and Scalable Optimization},
  year = {2025},
  url = {https://openreview.net/forum?id=tJqsZZBmmB}
}

@misc{upadhyaya2026autolyap,
  author = {Upadhyaya, Manu and Das Gupta, Shuvomoy and Taylor, Adrien B. and Banert, Sebastian and Giselsson, Pontus},
  title = {The {AutoLyap} software suite for computer-assisted {Lyapunov} analyses of first-order methods},
  year = {2026}, eprint = {2506.24076}, archivePrefix = {arXiv},
  primaryClass = {math.OC},
  url = {https://arxiv.org/abs/2506.24076v2},
  note = {{arXiv}:2506.24076, version 2 (2 March 2026); consulted 10 November 2026}
}

@misc{morrison2026sos,
  author = {Morrison, Kim},
  title = {{sos}: Sum-of-squares proofs in {Lean} 4},
  year = {2026},
  howpublished = {Software},
  note = {Revision fb7ae417609093f04cf0608dc92e9343550c2ae4. Accessed 10 November 2026},
  url = {https://github.com/leanprover/sos/tree/fb7ae417609093f04cf0608dc92e9343550c2ae4}
}

\clearpage
\appendix
\pdfbookmark[0]{Appendix}{appendix}
\section*{Appendix}
\section{Exact interface and declaration correspondence}\label{app:interface}
The parameters of the semantic and capture declarations are grouped below by role.
In the capture declaration, $p$ and $c$ denote the source binders \lean{policy} and \lean{reconstructed}. In both declarations, $x_\star$ denotes \lean{xStar}.
The proof call retains the actual \Lean argument order.

\subsection*{Type and structural parameters}
Here \lean{Type*} denotes an arbitrary universe.
The horizon is $T=n+1$.
\begin{interfaceparams}
\item $E:\mathtt{Type*}$
\item $[\mathtt{NormedAddCommGroup}\ E]$
\item $[\mathtt{InnerProductSpace}\ \R\ E]$
\item $[\mathtt{CompleteSpace}\ E]$
\item $n:\mathbb N$
\item $k:\mathbb N$
\item $s:\mathtt{SourceQ}\ n\ k$
\item $c:\mathtt{UpperCertQ}\ s.\mathtt{model}$
\item $t:\mathtt{TraceQ}\ n\ k$ \quad (capture interface only)
\item $p:\mathtt{PolicyQ}$ \quad (capture interface only)
\item $\mathtt{raw}:\mathtt{UpperCertQ}\ s.\mathtt{model}$ \quad (capture interface only)
\end{interfaceparams}

\subsection*{Optimization parameters}
\begin{interfaceparams}
\item $L:\R_{\ge0}$
\item $R:\R$
\item $f:E\to\R$
\item $h:E\to\R$
\item $g:E\to E$
\item $x_\star:E$
\item $x:\mathbb N\to E$
\end{interfaceparams}

\subsection*{Proof assumptions}
We abbreviate \lean{ProximalGradientScaling.IsSmoothProxRun} as \lean{IsSmoothProxRun}.
\begin{interfaceparams}
\item $\mathtt{hf}:\mathtt{IsSmoothConvexAnalytic}\ L\ f\ g$
\item $\mathtt{hh}:\mathtt{ConvexOn}\ \R\ \mathtt{Set.univ}\ h$
\item $\mathtt{hL}:0<(L:\R)$
\item $\mathtt{hR}:0\le R$
\item $\mathtt{hmin}:\forall y,\ f(x_\star)+h(x_\star)\le f(y)+h(y)$
\item $\mathtt{hrun}:\mathtt{IsSmoothProxRun}\ (n+1)\ L\ h\ g\ x$
\item $\mathtt{hradius}:\|x_0-x_\star\|^2\le R^2$
\end{interfaceparams}
Supply the acceptance premise for the chosen interface; the following entries are alternatives.
Both \Lean declarations name this premise \lean{hc}.
\begin{interfaceparams}
\item Semantic interface: $\mathtt{hc}:\mathtt{checkBoundQ}\ s\ c=\mathtt{true}$
\item Capture interface: $\mathtt{hc}:\mathtt{checkCapturedQ}\ s\ t\ p\ \mathtt{raw}\ c=\mathtt{true}$
\end{interfaceparams}
The conclusion of either declaration is \eqnref{eq:actual}.
The two-step reuse proof calls
\begin{quote}
\lean{ProximalGradientCaptureAudit.upper_analytic}\\
\lean{sourceQ traceQ policyQ rawUpperQ structuredUpperQ structured_check}\\
\lean{hf hh hL hR hmin hrun hradius}.
\end{quote}
The instance separately proves the source, trace, raw binding, direct policy, PSD, upper and captured checks. It also proves the exact bound and the policy count.
The companion map quotes the frozen reuse signature and the generated batch Replays.
Execution records cover the seven distinct batch Replays and the fresh-installation Replay.

\begingroup\small
\clubpenalty=110
\widowpenalty=110
\begin{longtable}{>{\raggedright\arraybackslash}p{.15\linewidth}>{\raggedright\arraybackslash}p{.37\linewidth}>{\raggedright\arraybackslash}p{.385\linewidth}}
\caption{Principal declaration families. Paths are under \lean{LeanPEP} unless otherwise stated; \lean{docs/evidence/claim-evidence.json} records full names, locations and source hashes.}\label{tab:map}\\
\toprule Claim & Declaration & File or evidence\\
\midrule\endfirsthead
\toprule Claim & Declaration & File or evidence\\
\midrule\endhead
Semantic entry & \lean{ActualFunction.ProximalGradientSparseAudit.upper_analytic} &
\lean{ActualFunction/ProximalGradientSparseAuditCore.lean}\\
Finite lifting & \lean{Adapters.ProximalGradientPerformance.rows_feasible_iff}, \lean{finite_upper_equivalence} &
\lean{Adapters/ProximalGradientPerformance.lean}\\
Capture entry & \lean{ActualFunction.ProximalGradientCaptureAudit.upper_analytic} &
\lean{ActualFunction/ProximalGradientCaptureAuditCore.lean}\\
Checker split & \lean{Adapters.ProximalGradientCapture.checkCapturedQ_eq_true_iff} &
\lean{Adapters/ProximalGradientCapture.lean}\\
Finite endpoint & \lean{ActualFunction.ProximalGradientCaptureAudit.finite_upper} &
Capture audit Core\\
Actual embedding & \lean{ActualFunction.ProximalGradient.actualData_constraints} &
\lean{ActualFunction/ProximalGradient.lean}\\
Minimizer bridge & \lean{ActualFunction.ProximalGradientScaling.certificate_upper_of_minimizer} &
\lean{ActualFunction/ProximalGradientMinimizerCore.lean}\\
PG exact pair & \lean{ActualFunction.PGOneCapturedTightness.finite_tightness}, \lean{actual_tightness} &
\lean{ActualFunction/PGOneCapturedTightness.lean}\\
GD exact pair & \lean{ActualFunction.GDOneCapturedTightness.finite_tightness}, \lean{actual_tightness} &
\lean{ActualFunction/GDOneCapturedTightness.lean}\\
Reuse instance & \lean{LeanPEP.Imported.ReuseR02.PG2.actual_upper} &
PG2 reuse case Replay\\
Target/row controls & \lean{Imported.CompositeRows.forged_local_passes}, \lean{forged_joint_rejected}, \lean{PGOne.bound_check}, \lean{one_rhs_preserved} &
\lean{Imported/CompositeRowsRegression.lean}\\
Batch instances & \lean{LeanPEP.Imported.Refine001.B01.actual_upper} through \lean{B08.actual_upper}, except B06 &
Seven batch Replays\\
Fresh installation & \lean{LeanPEP.Imported.Refine001.FreshInstall.actual_upper} &
Archived installation Replay\\
Other controls & \lean{LeanPEP.Imported.Reuse.Contrasts} (\cref{sec:controls}) &
Archived controls Replay\\
\bottomrule
\end{longtable}
\endgroup

For declarations beginning with \lean{ActualFunction}, \lean{Adapters} or \lean{Imported}, prepend \lean{LeanPEP.}. Shortened names share the preceding full namespace.
Timing, closure review and payload distinctness are supported by the corresponding execution records and audits.

\paragraph*{Additional declaration details.}
The structural binder \lean{SourceQ} abbreviates the fully qualified source type
\begin{quote}
\lean{LeanPEP.Adapters.ProximalGradientSparse.SourceQ}.
\end{quote}
The analytic bridge obtains first-order hypotheses through \lean{hf.toFirstOrder}.
Capture decomposition uses \lean{checkCapturedQ_eq_true_iff} and \lean{reconstructed_check}; the trace check in \tabref{tab:support} is \lean{checkTraceQ}.
In the GD finite route, feasibility includes \lean{RealizesModelQ}, \lean{CoreFeasible} and the model assumptions in addition to Gram PSD.
The one-step PG/GD results use \lean{analytic_witness_verified} and \lean{actual_tightness}.

\section{Generic implementation and reconstruction details}\label{app:engineering}
\begingroup\raggedright
The reusable entry imports the generic capture interface through
\[
\texttt{LeanPEP.ActualFunction.ProximalGradientCaptureAuditCore}.
\]
Its mathematical dependency graph contains normalization, scaling of PG runs and certificates, the minimizer bridge for arbitrary certificates, affine-vector evaluation, and source and representation checks.
Concrete applications and regressions are exercised by the full-library build and maintenance verification commands.
\par\endgroup

The generic closure contains 57 project modules, excluding external \toolname{mathlib} dependencies.
\lean{docs/pg-core-content-review.json} records each module's role, content hash and direct imports.
The consumer rejects missing, unreviewed or changed project sources.
Regression tests include concrete fixtures with both recognizable and renamed filenames to check that the guard treats reviewed contents consistently under these renamings.
The guard controls this engineering boundary; soundness follows from the mathematical theorems.

\subsection*{PG2 source and affine-solve details}
The native PG2 basis is $[x_\star,g_\star,x_0,g_0,s_1,g_1,s_2,g_2]$.
The PG2 source has twenty rows: a performance row, a radius row, twelve ordered smooth-pair rows and six ordered convex-pair rows.
The source validator checks the basis, row identities and order.
The consumer calls \lean{convert(...,2)} with the unit-scale \lean{leZero} trace convention.
Value-slot maps preserve both Gram axes and the original order of all source rows.

For the PG2 reuse input, the vector of unknowns in the affine system is $z=(b,\lambda_0,\ldots,\lambda_{19})$.
The radius multiplier has index 2 in the vector of unknowns. Its seed is rounded upward.
The pivot indices are $0,1,3,4,5,15,16$; the remaining fourteen seed entries are retained.
All indices here are zero based.
The nine equations are rechecked before forming the residual.

\subsection*{Exact comparisons for the reuse input}
The largest difference between the new and previous raw values occurs at multiplier 10:
\[
0.6714451237581617\ \longrightarrow\ 0.65794427613641, \qquad |\Delta|=\frac{135008476217517}{10000000000000000}.
\]
Each of the three historical payloads differs from the new raw input in 85 of the 158 covered entries.
This comparison measures the difference between raw solver endpoints.

The direct raw-to-candidate audit covers 158 entries, with maximum error
\[
\frac{49659234539500903}{49999900000000000000000} \simeq 9.93186677\times10^{-7}
\]
at multiplier 7. The maximum denominator among the covered candidate entries is 999998.
The accepted second grid yields six PSD factors and $B=62500/499999$.
The complete audit, selected data and generated proofs are linked by the artifact evidence index.

\tabref{tab:assistance} records the inputs used to construct each case's exact certificate.
\begin{table}[ht!]\centering\small
\setlength{\tabcolsep}{-0.3pt}
\begin{tabular}{p{.17\linewidth}p{.81\linewidth}}
Case & Information used by the untrusted construction
\\ \hline\hline
PG1 & Specific ordered multiplier family and exact rank-three PSD structure; coefficients inferred from raw within the declared budget.\\
PG2 and PG3 & Fixed source-affine systems, deterministic free-column grid seeds, residual reconstruction and exact factor checks; generation flag known-rate-used is false.\\
GD1 exact pair & Target $1/6$ and source/equality structure; generation flag known-rate-used is true.\\
All cases & Human choices of the profile and representation, informed by prior algorithms and classical rates.
\end{tabular}
\caption{Construction inputs and the generation flags recorded by each candidate's code path.}\label{tab:assistance}
\end{table}

\subsection*{Target and representation controls}
\begingroup\raggedright
In the namespace \lean{LeanPEP.Imported.CompositeRows} (file \lean{Imported/CompositeRowsRegression.lean}), the \lean{forgedSourceQ} construction has the local check \lean{forged_local_passes} and target rejection \lean{forged_joint_rejected}.
The synchronized Rows example uses the transformation rules \lean{LeanPEP.Imported.CompositeRows.PGOne.rulesQ}.
The declarations \lean{PGOne.rules_check}, \lean{PGOne.source_check}, \lean{PGOne.bound_check}, \lean{one_snapshot_roundtrip} and \lean{one_rhs_preserved} check the transformation and preservation of the certificate expression. All five declarations are in \lean{LeanPEP.Imported.CompositeRows}.
The supplementary controls are in \lean{LeanPEP.Imported.Reuse.Contrasts}. The target-valid looser certificate has the proved declaration \lean{looser_actual_upper}.
The evidence map records the original execution scope for these target and representation controls.
\par\endgroup

\section{Trust and reproduction}\label{app:reproduce}
\subsection{Trust and acceptance}\label{sec:trust}
The formal guarantee applies to the typed \Lean source, certificate and trajectory objects.
Exact certificate checks and the soundness theorems establish their mathematical relationships.
For a numerical capture, the source, trace and raw checks additionally connect the generated typed objects under the declared reconstruction policy.
The solver, capture program, JSON parser and candidate constructor supply untrusted data; exact parsing interprets each finite decimal literally.
Byte identities, conversion audits and execution logs make the correspondence between files and typed objects inspectable.
Parser correctness and cryptographic attestation of external computation remain outside the formal statement.
The accepted proofs use only \lean{propext}, \lean{Classical.choice} and \lean{Quot.sound}.

\subsection{Environment and installation}
The complete repository snapshot contains the subset required for the representative PG2 reproduction: the 57-module Core import closure, the PG2 consumer and its 21 required \toolname{Python} modules, the two archived input files, dependency locks and the software license.
The \lean{SOURCE_MANIFEST.json} manifest identifies these files within the complete code distribution.
The manuscript and its submission sources are distributed separately.

The offline consumer uses \toolname{Python}'s standard library.
Setup requires \toolname{Python} 3.10 or later, \toolname{Git}, \toolname{elan} and a supported C compiler. It also requires network access to the official upstreams and the matching \mathlib binary cache.
The toolchain is \Lean~4.31.0 with the nine revisions in \lean{lake-manifest.json}.
A clean matching external dependency cache can alternatively be supplied with \texttt{--cache}; project \Lean objects are excluded.
The setup and consumer entry points both clear inherited \Lean and \toolname{Python} search paths.
Verification of the supplied input requires no solver installation or new numerical capture.

\subsection{Representative verification and outputs}\label{app:review-run}
From the root of a fresh checkout or extracted archive of the commit specified in \appref{app:artifacts}, set \lean{OUT} to an unused absolute directory outside the source tree.
With the prerequisites above, run
\begin{small}\begin{verbatim}
  python3 scripts/setup.py --output "$OUT/setup"
  python3 tools/replay_pg_certificate.py verify \
    --source-root "$PWD" \
    --input "$PWD/data/cases/pg/two-step-reuse/capture.json" \
    --candidate "$PWD/data/cases/pg/two-step-reuse/selected_exact.json" \
    --output "$OUT/pg2" --namespace LeanPEP.Imported.Review.PG2
\end{verbatim}\end{small}
Setup verifies the manifest and lockfile, obtains the dependencies and matching binary cache, then builds Core from the supplied sources.
Verification consumes the archived input offline.
On success, both commands exit with code zero. The \lean{setup/result.json} file reports \lean{core_built}. The \lean{pg2/result.json} file reports \lean{passed}, \lean{actual_upper_verified=true} and bound $62500/499999$.
The pg2 directory also contains \lean{Replay.lean}, \lean{replay.log}, the direct audit, input identity and selected exact data.
The generated theorem is \lean{LeanPEP.Imported.Review.PG2.actual_upper}, with the full function and run contract of \cref{sec:contract}. The generated proof discharges the certificate acceptance premise by computation.

A separate, focused review distribution containing the required subset was independently unpacked and verified using the locked external-cache branch. Core was rebuilt from source during that verification.
Core and Replay both exited with code zero. The printed proof dependencies contained only the standard axioms listed above.
That execution record is scoped to the focused distribution and retains the environment, exact commands, cache provenance and logs.

\subsection{Additional maintained verification modes}
The complete repository snapshot also provides \lean{docs/quickstart.md} and \lean{scripts/verify.py}.
Its \texttt{quick} mode builds Core, runs interface and consumer tests, and verifies the same supplied PG2 certificate.
The \texttt{full} mode additionally repeats recovery, compares selected bytes, and runs the complete maintenance plan. The \texttt{tightness} mode exposes both exact one-step pairs.
Each mode supports \texttt{--list} to inspect its plan before execution.

The direct consumer also supports \texttt{recover} to construct and check a candidate with the fixed strategy. Omit \texttt{--candidate} when using this mode.
Its generated checks and analytic theorem are the same as in verification mode.
\appref{app:artifacts} indexes historical source, batch and installation evidence by the original source identities.

\section{Artifact identities and evidence}\label{app:artifacts}
The code snapshot at \url{https://github.com/weartem13/lean-pep} is identified by commit \href{https://github.com/weartem13/lean-pep/commit/b4a2467eb254d891a5f29979555523ffdfb77b2e}{\texttt{b4a2467eb254d891a5f29979555523ffdfb77b2e}}.
It combines implementation, documentation, certificates and execution evidence.
The manuscript is distributed separately.

The repository's \lean{data/paper/evidence-index.json} manifest records the source used for the experiments in \cref{sec:batch} and \cref{sec:cost}. The \lean{SOURCE_MANIFEST.json} manifest identifies the maintained source tree. The authors separately maintain \lean{paper-manifest.json} as an internal record of manuscript identity.
The results retain the identities of the sources used for their original executions, as recorded in the evidence index.

The evidence index connects every retained experimental input, selected candidate, Replay and execution record to its original path and hash.
The bytes of raw captures, selected certificates and generated \Lean proofs remain unchanged.
In public copies of historical logs, personal absolute-path prefixes have been replaced. The index gives original and derived identities and the scope of these replacements.
Original records remain archived.

The case directories distinguish PG1 exact, PG2 initial/reuse, PG3, GD1 gap/exact and GD2 evidence.
\lean{data/paper/pg2-batch} contains all eight configurations and captures, the seven distinct accepted candidates and Replays, all returned grid diagnostics, exact deduplication relationships and execution costs.
\lean{data/paper/installation} contains the fresh-installation records.
The compact \lean{docs/evidence/claim-evidence.json} connects the main statements to declarations and these accessible records.

The evidence records 1163 \toolname{Python} tests for the frozen source. It also records twenty targeted tests for the refinement: six harness tests, four outer-entry tests and ten consumer tests.
Build, smoke, Replay, batch, installation and profiling results are indexed by their corresponding source and execution.
The maintained artifact provides its own verification plan, executable entry points and corresponding result records.

\begin{center}\small\begin{tabular}{lr}
Reuse input object & Bytes\\ \hline
Raw capture JSON & 140264\\
Selected exact JSON & 325280\\
Generated Replay source & 106850
\end{tabular}\end{center}
The 158 policy entries contain 83 nonzero rationals and have maximum denominator bit length 20.
The 54 separately scoped rationals for factor weights and vectors have a maximum denominator bit length of 106.
The factor policy checks factor coefficients. The coefficient denominator cap does not apply to these coefficients.
The table reports generated data sizes and the size of the Replay source that instantiates the generic theorem.

\end{document}